# Enhanced critical current density in the pressure-induced magnetic state of the high-temperature superconductor FeSe

Soon-Gil Jung[1,*], Ji-Hoon Kang[1], Eunsung Park[1], Sangyun Lee[1], Jiunn-Yuan Lin[2], Dmitriy A. Chareev[3], Alexander N. Vasiliev[4,5,6], and Tuson Park[1,*]

[1]*Department of Physics, Sungkyunkwan University, Suwon 440-746, Republic of Korea*

[2]*Institute of Physics, National Chiao Tung University, Hsinchu 30010, Taiwan*

[3]*Institute of Experimental Mineralogy, Russian Academy of Sciences, Chernogolovka, Moscow Region 142432, Russia*

[4]*Low Temperature Physics and Superconductivity Department, Physics Faculty, Moscow State University, 119991 Moscow, Russia*

[5]*Theoretical Physics and Applied Mathematics Department, Institute of Physics and Technology, Ural Federal University, Ekaterinburg 620002, Russia*

[6]*National University of Science and Technology "MISiS", Moscow 119049, Russia*

*Tuson Park (e-mail: tp8701@skku.edu), *Soon-Gil Jung (e-mail: prosgjung@gmail.com)

**We investigate the relation of the critical current density ($J_c$) and the remarkably increased superconducting transition temperature ($T_c$) for the FeSe single crystals under pressures up to 2.43 GPa, where the $T_c$ is increased by ~8 K/GPa. The critical current density corresponding to the free flux flow is monotonically enhanced by pressure which is due to the increase in $T_c$, whereas the depinning critical current density at which the vortex starts** to move is more influenced by the pressure-induced magnetic state compared to the increase of $T_c$. Unlike other high-$T_c$ superconductors, FeSe is not magnetic, but superconducting at ambient pressure. Above a critical pressure where magnetic state is induced and coexists with superconductivity, the depinning $J_c$ abruptly increases even though the increase of the zero-resistivity $T_c$ is negligible, directly indicating that the flux pinning property compared to the $T_c$ enhancement is a more crucial factor for an achievement of a large $J_c$. In addition, the sharp increase in $J_c$ in the coexisting superconducting phase of FeSe demonstrates that vortices can be effectively trapped by the competing antiferromagnetic order, even though its antagonistic nature against superconductivity is well documented. These results provide new guidance toward technological applications of high-temperature superconductors.

The technological application of superconductors hinges on how to preserve a zero-resistance state at high temperature while maintaining large electrical currents. The discovery of copper-based high-temperature superconductors (HTSs) brought great excitement not only because of its unconventional superconducting nature, but also because of its high superconducting transition temperature ($T_c$), which was expected to open the door for revolutionary applications at temperatures higher than liquid nitrogen temperature (=77 K) (refs. 1-3). A key issue for practical applications of superconductors is the necessity to increase the value of the depinning critical current density ($J_c$), at which magnetic flux lines (or vortices) start to flow and energy dissipation occurs. For decades, several approaches effectively enhanced the $J_c$ of HTSs by introducing and/or manipulating the extrinsic defects that suppress superconductivity[4,5]. Because the flux lines have a normal state within the core, they tend to be pinned at defects where superconductivity is suppressed, i.e., extrinsic pinning effects.

Another possible approach to improve the $J_c$ is associated with an intrinsic property of materials, e.g., a coexisting order with superconductivity as an intrinsic pinning source. Recently, it has been proposed that magnetism may be conducive to holding the vortex, which leads to the enhancement of the $J_c$ (refs. 6-11). Several high-$T_c$ superconductors, such as $La_{2-x}Sr_xCuO_4$ and $Ba(Fe_{1-x}Co_x)_2As_2$, are candidate materials for the intrinsic pinning because superconductivity occurs in the vicinity of an antiferromagnetically ordered state[6-8]. Superconductivity in those materials, however, requires a chemical substitution that inherently induces defects or site disorder, intertwining the effects of impurities and intrinsic pinning on $J_c$. In addition, it is still controversial if the magnetic order arises from macroscopically phase separated domains or from an intrinsic coexisting phase on a microscopic level. Therefore, in order to clarify the role of the intrinsic pinning on $J_c$, it is crucial to perform a systematic study on a high-$T_c$ compound that is superconducting in stoichiometric form and tunable between superconducting and magnetic ground states by non-thermal control parameters.

The binary high-$T_c$ superconductor FeSe is a promising candidate to probe the effects of the intrinsic pinning and the $T_c$ on the $J_c$, because superconductivity which appears at ~ 10 K without introducing a hole or electron in the parent compound is greatly tunable up to 37 K by application of pressure[12,13]. In addition, an emergence of magnetic state at pressure ~ 0.8 GPa makes it a more interesting material in its basic properties and application issues[14,15]. A $T_c$ above 100 K in FeSe monolayer shows its promising potential for the possibility of application[16]. In the following, we report the evolution of the critical current density ($J_c$) of FeSe single crystals as a function of pressure in connection with the increase of $T_c$.

The current-voltage ($I-V$) characteristic curves as well as temperature dependences of the electrical resistivity show a sharp contrast across the critical pressure ($P_c$=0.8 GPa) above which $\mu$SR measurements reported a pressure-induced AFM state that coexists with superconductivity[14,15]. There are a few interesting behaviours. First, the superconducting (SC) transition is sharp at low pressures, but becomes broader in the coexisting SC state for $P > P_c$. Secondly, temperature dependence of the critical current density follows the prediction by the $\delta T_c$-pinning at low pressures ($P < P_c$), while the $\delta l$-pinning becomes more effective at higher pressures. Thirdly, amplitude of $J_c$ is strongly enhanced in the coexisting state. The fact that physical pressure does not induce extra disorder suggests that the enhancement in $J_c$ as well as the change in the pinning mechanism in the coexisting phase arises from the antiferromagnetically ordered state.

# Results

Figures 1(a) and (b) representatively shows the in-plane electrical resistivity ($\rho_{ab}$) of FeSe as a function of temperature for several pressures. For clarity, $\rho_{ab}(T)$ for different pressures were rigidly shifted upwards. At ambient pressure, a change in the slope of $\rho_{ab}$ occurs at 75 K due to the tetragonal to orthorhombic structural phase transition. Unlike other iron-based superconductors, this structural transition is not accompanied by a magnetic phase transition. The structural transition temperature ($T_s$), which is assigned as a dip in $d\rho_{ab}/dT$, progressively decreases with increasing pressure at a rate of -36.7 K/GPa and is not observable for pressures above 1.3 GPa where $T_s$ becomes equal to the superconducting transition temperature $T_c$, as shown in Fig. 1(d). With further increasing pressure, an additional feature appears in the normal state as a dip or a slope change in $d\rho_{ab}/dT$, as shown in Fig. 1(e). In contrast to $T_s$, this new characteristic temperature linearly increases with pressure and is nicely overlaid with the $T_N$ determined from $\mu$SR results[14], showing that the resistivity anomaly arises from the paramagnetic to antiferromagnetic phase transition, as described in Fig. 1(f).

Figure 1(c) presents that the temperature for the onset of the superconducting transition ($T_{c,on}$) gradually increases with increasing pressure at a rate of 8 K/GPa. Also, the transition width $\Delta T_c$, which was defined as the difference between the 90 and 10% resistivity values of the normal state at $T_{c,on}$, decreases with increasing pressure at low pressures because of the enhanced superconductivity under pressure. At pressures $P > 0.8$ GPa, where superconductivity coexists with a magnetically ordered state on a microscopic scale[14,15], $\Delta T_c$ becomes broader even though $T_{c,on}$ increases with increasing pressure. The dichotomy between $T_{c,on}$ and $\Delta T_c$ in the coexisting phase suggests that the pressure-induced antiferromagnetic phase acts as an additional source for breaking Cooper pairs.

Correlation between the anomalous broadening in the $\Delta T_c$ and the magnetic phase is further supported by a qualitative difference in the current-voltage ($I-V$) curves of FeSe across the critical pressure $P_c$. As shown in Figs. 2(a)-(d), the voltage curve sharply decreases with decreasing current at 0.41 GPa, i.e., the pressure where superconductivity itself only exists. In the coexisting phase ($P > P_c$), on the other hand, the voltage curve develops a knee with decreasing current. Figure 2(d) summarizes pressure evolution of the transition broadening in the $I-V$ curve at 7 K. These anomalous broadenings in the $I-V$ curves are also considered due to the pressure-induced antiferromagnetic state.

## Discussion

Two characteristic critical currents, $I_c$ and $I_f$ from the $I-V$ curves, are marked by the two arrows in Fig. 2(d). The depinning critical current ($I_c$) was obtained from the 1 $\mu$V criterion where the vortices start to move and the free-flux-flow (FFF) current ($I_f$) was obtained from the point where vortices are no longer affected by the pinning sites and therefore move freely[10,17]. Figures 3(a) and (b) describes the temperature dependence of the critical current densities $J_f$ and $J_c$ estimated from $I_f$ and $I_c$, respectively. Both $J_f$ and $J_c$ were significantly improved with increasing pressure. The FFF current density $J_f(T)$, which is concerned with thermally activated flux flow with increasing $T_{c,on}$, is best explained by the empirical relation $J_f(T) \sim [1-(T/T_{c,on})^n]$, with $n = 2.6\pm0.2$ indicated by solid lines in Fig. 3(a). The curves all collapse onto a single curve, as shown in Fig. 3(c), which cannot be explained by the depairing current density ($J_d$) given by $J_d(t) \propto (1-t^2)^{3/2}(1+t^2)^{1/2}$ (dashed line)[18], nor by the Joule heating, $J_{heating}$ ($\Delta T \propto J^2$) which is caused by the contact resistance (dotted line)[19]. Rather, they collapse onto the curve expected from the $\delta T_c$-pinning mechanism (solid line), $J_f(t) \propto (1-t^2)^{7/6}(1+t^2)^{5/6}$, suggesting that the temperature dependence of the FFF current density is primarily determined by spatial variations in $T_c$ (refs. 20, 21).

Figure 3(b) shows the pressure evolution of the depinning critical current density ($J_c$), usually called the critical current density, as a function of temperature. At 1.8 K, the lowest temperature measured, $J_c$ increases in commensurate with $T_{c,0}$ with increasing pressure, while $J_c$ in the coexisting phase is strongly enhanced from 1.89 kA/cm$^2$ at 0.41 GPa (red circles) to 3.24 kA/cm$^2$ at 1.22 GPa (blue triangles). Here, we used the zero-resistivity SC transition temperature ($T_{c,0}$) with applied current density ($J$) ~ 1 A/cm$^2$. Resistance is not zero any more above the $J_c$ where vortices start to move, which is significantly influenced on the pinning properties of samples, such as pinning strength, density of pinning sites, and so on. Therefore, the $J_c$ comparison by the $T_{c,0}$ is reasonable than the comparison by the $T_{c,on}$. Considering that the increase in $T_{c,0}$ is negligible at 1.2 GPa, the anomalous jump in $J_c$ as shown in Fig. S1 in SI, deviates from the trend in $J_c$ as a function of $T_{c,0}$, underlining that an additional source of pinning is indeed required to explain this anomaly. The possibility of the enhancement in $J_c$ due to improved grain boundary connectivity has been reported in some high-$T_c$ cuprate superconductors[22,23] or in the iron-based polycrystalline superconductor Sr$_4$V$_2$O$_6$Fe$_2$As$_2$ (ref. 24). Because the studied FeSe samples are single crystalline specimens, however, the lack of a weak-link behaviour in the field dependence of $J_c$ rules out the possibility of grain boundary as the additional pinning source (see Fig. S2 in SI). Rather, the simultaneous enhancement in $J_c$ and appearance of antiferromagnetism indicate that the pressure-induced magnetic state leads to an inhomogeneous SC phase and is conducive to the trapping of magnetic flux lines. With further increase in pressure, both $J_c$ and $T_{c,0}$ increase.

The additional flux pinning caused by the antiferromagnetic (AFM) order in the FeSe is reflected in the different temperature dependence of $J_c$ across the critical pressure $P_c$. As shown in Fig. 3(d), the normalized self-field critical current density $J_c(t_0)$ as a function of the reduced temperature ($t_0=T/T_{c,0}$) is well described by the $\delta T_c$-pinning mechanism (solid line) for $P < P_c$, where the $T_c$ fluctuates due to defects, such as Se deficiencies and point defects, which are the main sources for trapping the vortices. For $P \geq P_c$, however, $J_c(t_0)$ shows a completely different behaviour: the curvature of $J_c$ near $T_{c,0}$ is positive, while it is negative at lower pressures. Also with increasing pressure, $J_c$ deviates further away from the $\delta T_c$-pinning and at 2.43 GPa becomes close to the curve predicted by $\delta l$-pinning (dashed line), $J_c(t) \propto (1-t^2)^{5/2}(1+t^2)^{-1/2}$, suggesting that spatial fluctuations in the mean free path ($l$) of the charge carrier becomes important for flux pinning at high pressures[21]. As shown in Fig. S3 in SI, the pressure-induced crossover in $J_c(T)$ is almost independent of the magnetic field, indicating that the vortex pinning within the AFM phase is robust against variations in the magnetic field strength.

A similar crossover from $\delta T_c$-pinning to $\delta l$-pinning has been reported in MgB$_2$ when additional pinning sources, such as grain boundaries or inclusions of nanoparticles by chemical doping, were introduced[25] or hydrostatic pressure was applied[26]. In the present study, a broadening of superconducting transition with the pressure-induced AFM state is important for the crossover. A possibility of enhanced mean free path ($l \propto \xi$) fluctuations due to the competition between superconducting and AFM order parameters and change in the superconducting coherence length ($\xi$) with pressure may be closely related to the crossover because the disorder parameter that characterizes the collective vortex pinning properties is proportional to $\xi$ and to $1/\xi^3$ for $\delta T_c$- and $\delta l$-pinning, respectively[21,26]. As shown in Fig. S4 in SI, the values of the upper critical field $H_{c2}(0)$ increase with applied pressure, indicating that the change in $\xi$ may be of some relevance to the crossover.

Figure 4(a) shows a contour plot of the free-flux-flow current density ($J_f$) for FeSe as a function of temperature and pressure at zero field, where the colours represent different values of $J_f$. Also plotted are the structural and the magnetic phase boundaries that are obtained from the electrical resistivity measurements; these boundaries are consistent with those reported in previous works[14,15]. The contour of $J_f$ monotonically increases with an increase in $T_c$ by pressure, while $J_c$ deviates from the monotonic pressure evolution of $J_f$. Instead, the contour of $J_c$ reflects the appearance of the pressure-induced AFM phase, as shown in Fig. 4 (b). The $J_c$ as well as the $T_{c,0}$ gradually increases with increasing pressure, however near the critical pressure where AFM phase is induced, $J_c$ shows a high increase compared to $T_{c,0}$, as mentioned in Fig. 3(b). We note that $J_c$ shows a dome shape centred around 2.1 GPa, the projected critical pressure where the tetragonal to orthorhombic structural phase transition temperature is extrapolated to $T = 0$ K inside the dome of superconductivity[27]. A Possibility of flux pinning by structure transition had been reported in the superconducting A15 compounds such as $V_3Si$ (refs. 28,29), and further work is in progress to better understand the role of structural fluctuations in producing the peak in $J_c$.

## Conclusions

In conclusion, we studied the correlation between superconducting transition temperature and critical current density for the high-$T_c$ superconductor FeSe. Both $T_{c,on}$ and $J_f$ increase with pressure, which is insensitive to the presence of AFM states, on the other hand, the superconducting transition width becomes considerably broader with the emergence of the AFM phase, and $J_c$ is prominently enhanced in the coexisting phase. This behaviour reflects that the AFM phase not only provides an additional source of vortex pinning, but also makes the system susceptible to the inhomogeneous SC phase. Even though these observations are only specific to FeSe, they are expected to guide theoretical as well as experimental efforts to better understand the vortex pinning in the high-$T_c$ superconductors where competing orders coexist on a microscopic scale. Further, when combined with well-known extrinsic pinning techniques, intrinsic magnetic pinning will provide a blueprint for greatly enhancing the critical current density, thereby bringing one step closer to the technological applications of high-temperature superconductors.

**Methods**

The $c$-axis-oriented high-quality FeSe$_{1-\delta}$ ($\delta$=0.04±0.02) single crystals with a tetragonal structure (space group P4/*nmm*) were synthesised in evacuated quartz tubes in permanent gradient of temperature by using an AlCl$_3$/KCl flux. The synthesis technique used to fabricate the FeSe single crystals and their high-quality are described in detail elsewhere[30,31]. The current-voltage ($I-V$) characteristics of FeSe were measured under hydrostatic pressures of 0.00, 0.41, 1.22, 1.72, 2.00, and 2.43 GPa. The physical pressure was applied by using a hybrid clamp-type pressure cell with Daphne 7373 as the pressure-transmitting medium, and the value of the pressure at low temperatures was determined by monitoring the shift in the $T_c$ of high-purity lead (Pb) as a manometer. The $I-V$ characteristic measurements under pressure were performed in the physical property measurement system (PPMS 9T, Quantum Design), where the electrical current was generated by using an Advantest R6142 unit and the voltage was measured by using an HP34420A nanovoltmeter. The depinning critical current ($I_c$) was obtained from the 1 $\mu$V criterion instead of 1 $\mu$V/cm in the $I-V$ curves due to a small size of FeSe single crystals[10,32]. A few layers of FeSe in the FeSe single crystals were easily exfoliated by using adhesive tape, which is similar to the exfoliation technique that is used for graphene[33]. The size of the measured crystals is typically 590 × 210 × 5 $\mu$m$^3$. Quasi-hydrostatic pressure was achieved by using a clamp-type piston-cylinder pressure cell with Daphne oil 7373 as the pressure-transmitting medium. The magnetic fields were applied parallel ($H//ab$) to the *ab*-plane of the samples.

**Acknowledgments**

We thank W. N. Kang, S. Lin, and J. D. Thompson for helpful discussions. This work was supported by the National Research Foundation (NRF) of Korea grant funded by the Korean Ministry of Science, ICT and Planning (No. 2012R1A3A2048816). This work was supported in part by the Ministry of Education and Science of the Russian Federation within the framework of the Increase Competitiveness Program of the National University of Science and Technology "MISiS" (Contract No. K2-2014-036).

**Figure legends**

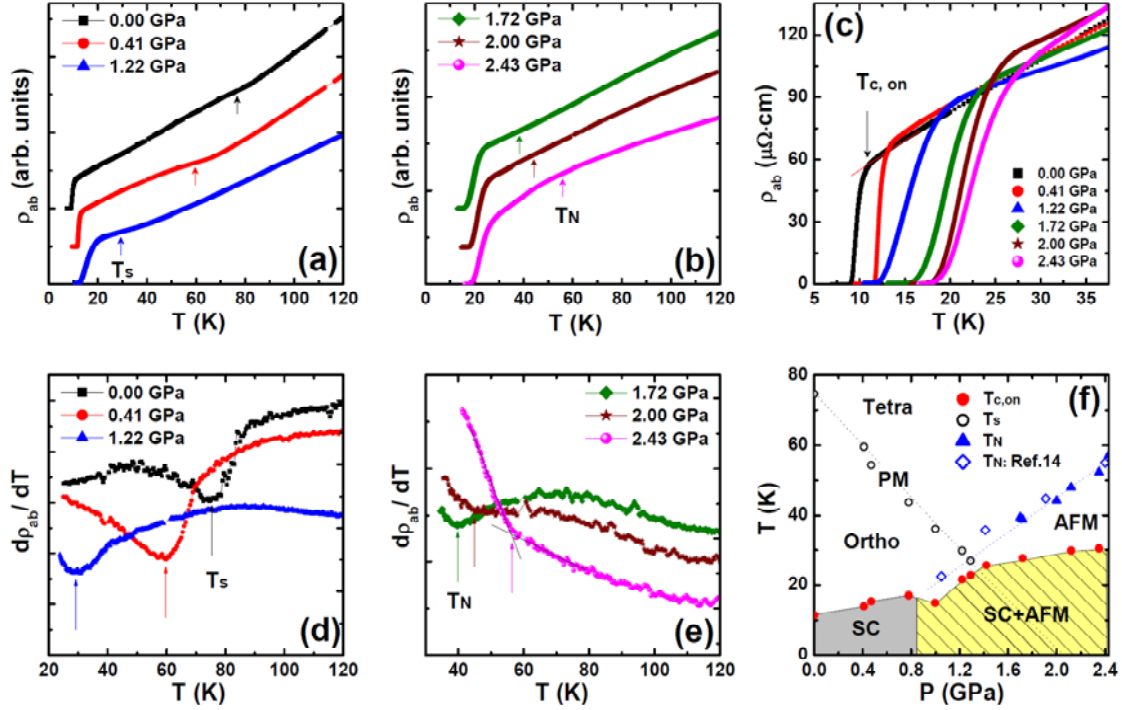

**Figure 1 Electrical resistivity and phase diagram of FeSe single crystals.** (a) and (b) In-plane electrical resistivity ($\rho_{ab}$) is plotted as a function of temperature for selective pressure. Arrows mark the structural ($T_s$) and antiferromagnetic phase transition ($T_N$) in (a) and (b), respectively. $\rho_{ab}$ for different pressures were rigidly shifted upwards for clarity. (c) $\rho_{ab}$ is magnified near the superconducting transition region, where $T_{c,on}$ is defined as the onset temperature of the SC phase transition. (d) and (e) First temperature derivative of the resistivity is shown as a function of temperature. Arrows mark $T_s$ and $T_N$ in (d) and (e), respectively. (f) Temperature-pressure phase diagram of FeSe. SC, AFM, and PM stand for superconducting, antiferromagnetic, and paramagnetic phase, respectively. Tetra and Ortho are the acronym of tetragonal and orthorhombic crystal structure.

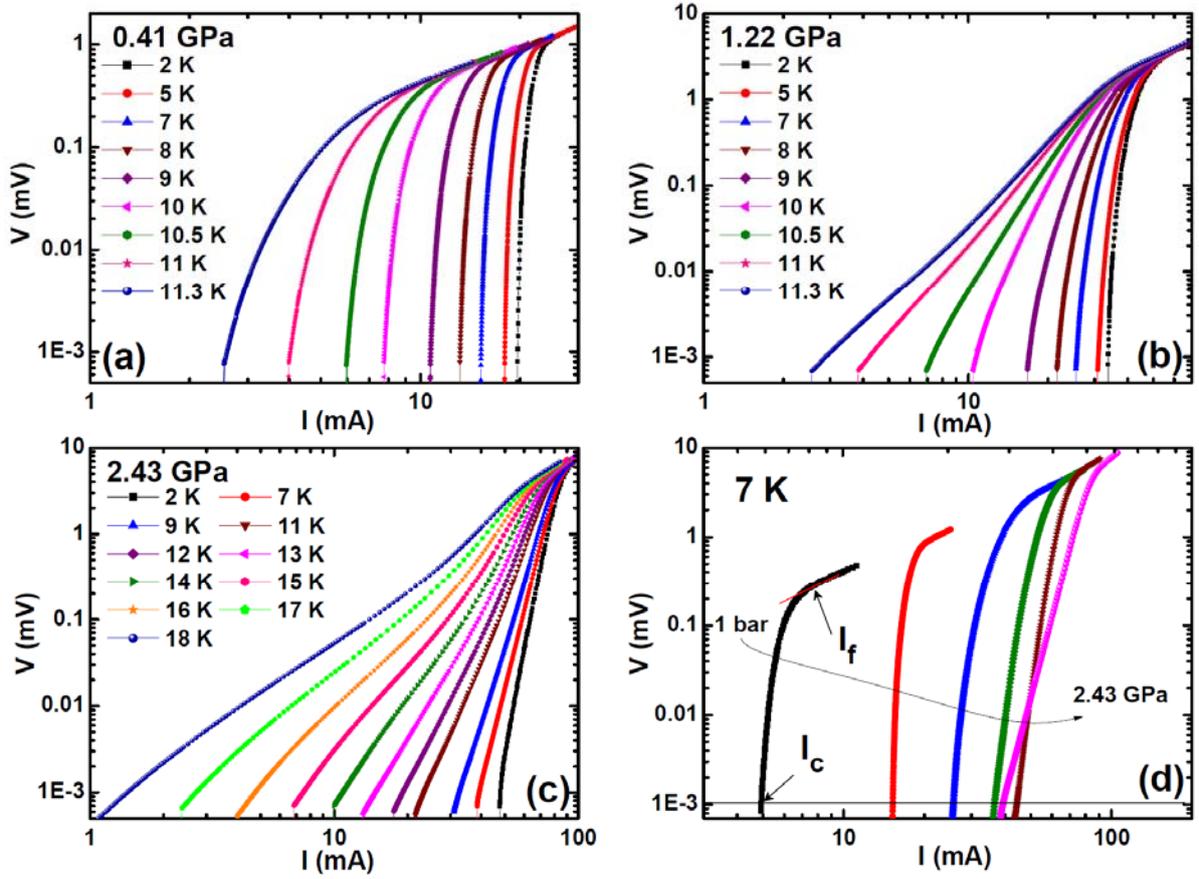

**Figure 2 Evolution of transport properties of FeSe single crystals under pressure**. (a)-(c) Logarithmic plots of the current-voltage ($I-V$) results at pressures of 0.41, 1.22 and 2.43 GPa. The $I-V$ curves become broader at pressure above 0.8 GPa, where an AFM phase is induced. (d) Pressure evolution of the isothermal $I-V$ curves at 7 K. The depinning critical current ($I_c$) is estimated by using the criterion of 1 $\mu$V, and the free-flux-flow critical current ($I_f$) is the value of the current at the inflection point, both of which are denoted by arrows.

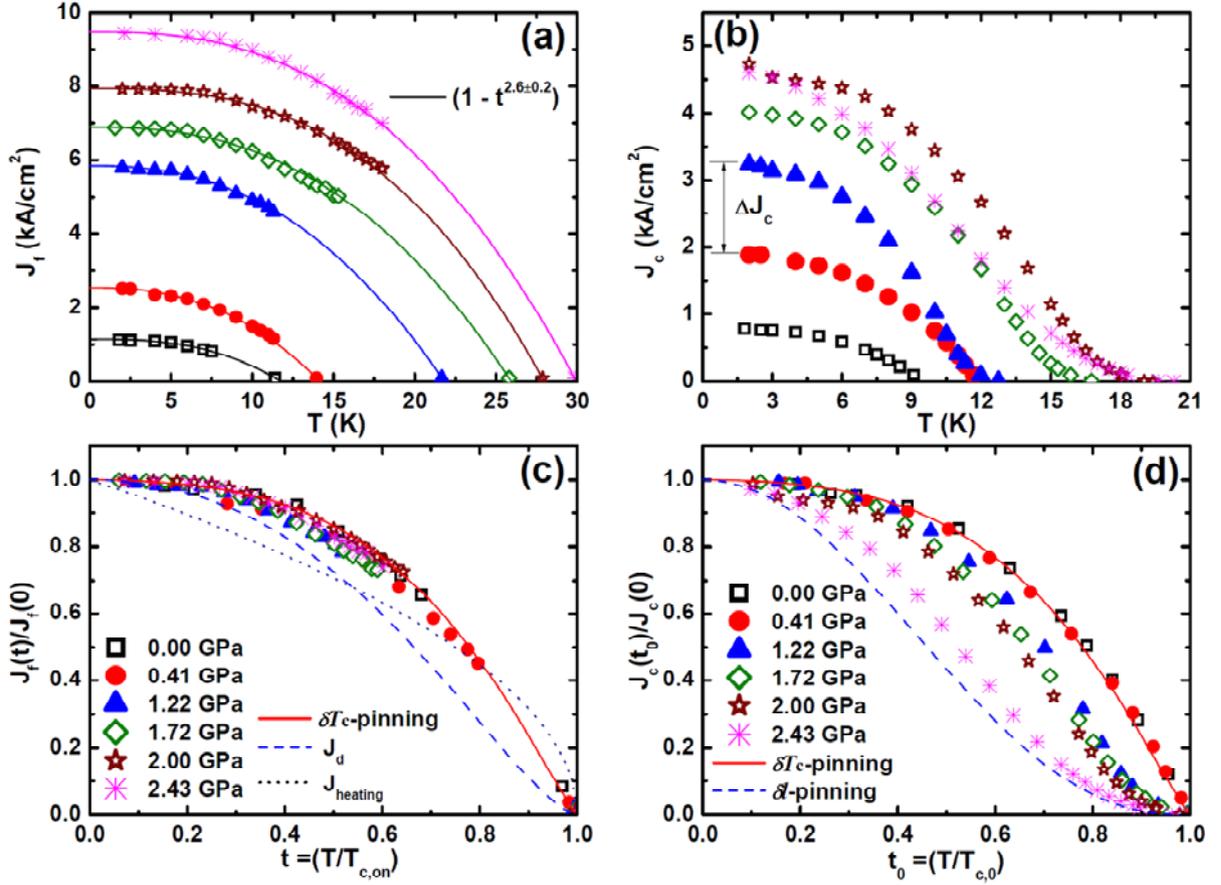

**Figure 3 Critical current densities of FeSe and the flux pinning mechanism under pressure.** (a) and (b), The free-flux-flow critical current density $J_f(T)$ monotonically increases with increasing $T_{c,on}$ by pressure and is well explained by the relation $1-(T/T_{c,on})^{2.6\pm0.2}$ over the entire pressure ranges (solid lines). On the other hand, the depinning critical current density $J_c(T)$ reveals a large enhancement at 1.22 GPa (solid triangles) even though the $T_{c,0}$ is similar to the value at 0.41 GPa (solid circles). $\Delta J_c$ is the jump in the critical current density at 1.22 GPa, which accounts for about 70% increase from that at 0.41 GPa. (c) Normalized $J_f(t)$ is plotted as a function of the reduced temperature $t$ (= $T/T_{c,on}$) for several pressures. All the curves collapse together, indicating that the underlying mechanism for the $J_f$ is independent of enhanced $T_{c,on}$ by pressure. The $J_f(t)$ curves follow the prediction by $\delta T_c$-pinning (solid line) – see the text for detailed discussion. (d) Normalized $J_c(t_0)$ is plotted as a function of another reduced temperature $t_0$ (=$T/T_{c,0}$), where $T_{c,0}$ is the zero-resistance transition temperature. $J_c(t_0)$ closely follows the prediction from $\delta T_c$-pinning at low pressures, while it deviates from $\delta T_c$-pinning at pressures above a critical pressure (=0.8 GPa), above which a magnetic state is induced. With further increase in pressure, $J_c(t_0)$ crosses into a region where $\delta l$-pinning dominates its temperature dependence.

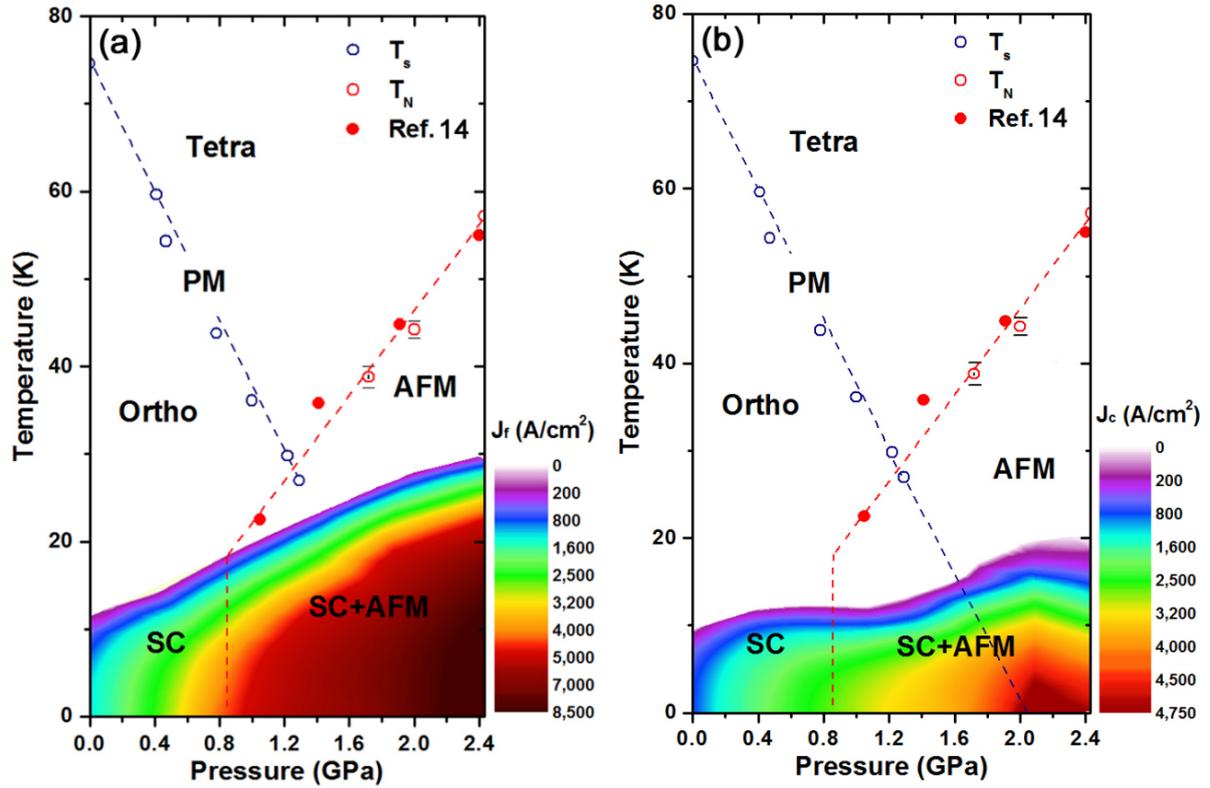

**Figure 4 Phase diagram of the critical current densities, $J_f$ and $J_c$.** (a) The free-flux-flow critical current density ($J_f$), above which vortices flow freely, is plotted as a function of temperature and pressure. Here, the colour represents the absolute value of $J_f$. The magnetic and the superconducting (SC) transition temperatures based on the resistivity measurements are also plotted. For reference, we show the phase transition temperature from paramagnetic (PM) to antiferromagnetic (AFM) states based on the $\mu$SR measurements in ref. 14 (solid red circles). (b) A contour map of the depinning critical current density ($J_c$) is plotted as a function of temperature and pressure, where the colour represents the absolute value of $J_c$.

# Supplementary Information for "Enhanced critical current density in the pressure-induced magnetic state of the high-temperature superconductor FeSe"

Soon-Gil Jung[1,*], Ji-Hoon Kang[1], Eunsung Park[1], Sangyun Lee[1], Jiunn-Yuan Lin[2], Dmitriy A. Chareev[3], Alexander N. Vasiliev[4,5,6], and Tuson Park[1,*]

[1]*Department of Physics, Sungkyunkwan University, Suwon 440-746, Republic of Korea*

[2]*Institute of Physics, National Chiao Tung University, Hsinchu 30010, Taiwan*

[3]*Institute of Experimental Mineralogy, Russian Academy of Sciences, Chernogolovka, Moscow Region 142432, Russia*

[4]*Low Temperature Physics and Superconductivity Department, Physics Faculty, Moscow State University, 119991 Moscow, Russia*

[5]*Theoretical Physics and Applied Mathematics Department, Institute of Physics and Technology, Ural Federal University, Ekaterinburg 620002, Russia*

[6]*National University of Science and Technology "MISiS", Moscow 119049, Russia*

**In this supplement, we describe methods and show additional data and analyses that support the results in the main text.**

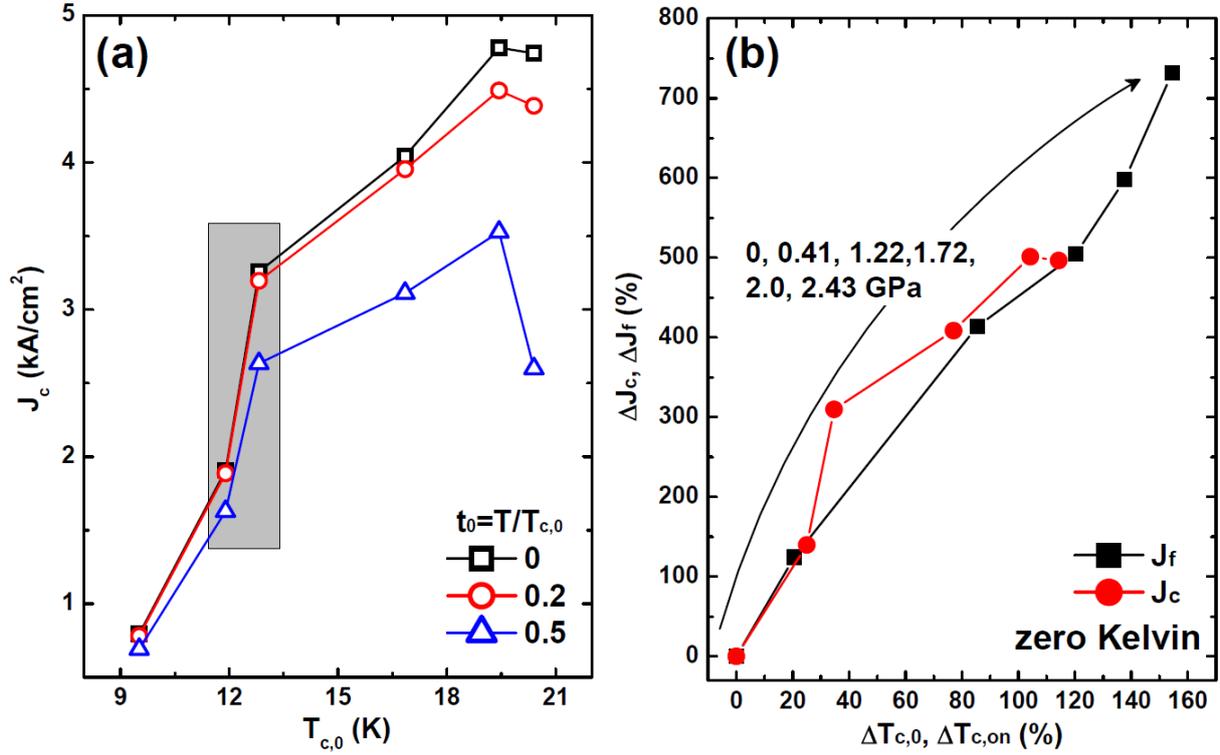

**Figure S1.** (a) The depinning critical current density $J_c$ as a function of $T_{c,0}$ for reduced temperatures ($t_0=T/T_{c,0}$) of 0, 0.2 and 0.5. Here, $T_{c,0}$ is the zero-resistance temperature. Appearance of the sudden jump in $J_c$ at 1.22 GPa, which is highlighted by the shade, is considered as due to magnetic pinning induced by the AFM phase. (b) Critical current densities of $\Delta J_f$ and $\Delta J_c$ as functions of $\Delta T_{c,on}$ and $\Delta T_{c,0}$, respectively, where $\Delta J_f = [J_f(P)-J_f(0)]/J_f(0)\times 100\%$, $\Delta J_c = [J_c(P)-J_c(0)]/J_c(0)\times 100\%$, $\Delta T_{c,on} = [T_{c,on}(P)-T_{c,on}(0)]/T_{c,on}(0)\times 100\%]$, $\Delta T_{c,0} = [T_{c,0}(P)-T_{c,0}(0)]/T_{c,0}(0)\times 100\%$, and $J_f(0)$, $J_c(0)$, $T_{c,on}(0)$, $T_{c,0}(0)$ are critical current densities and critical temperatures at ambient pressure. The $J_f$ shows a monotonic increase with increase in $T_{c,on}$, on the other hand, $J_c$ at 1.22 GPa shows a large enhancement despite a slight increase of $T_{c,0}$.

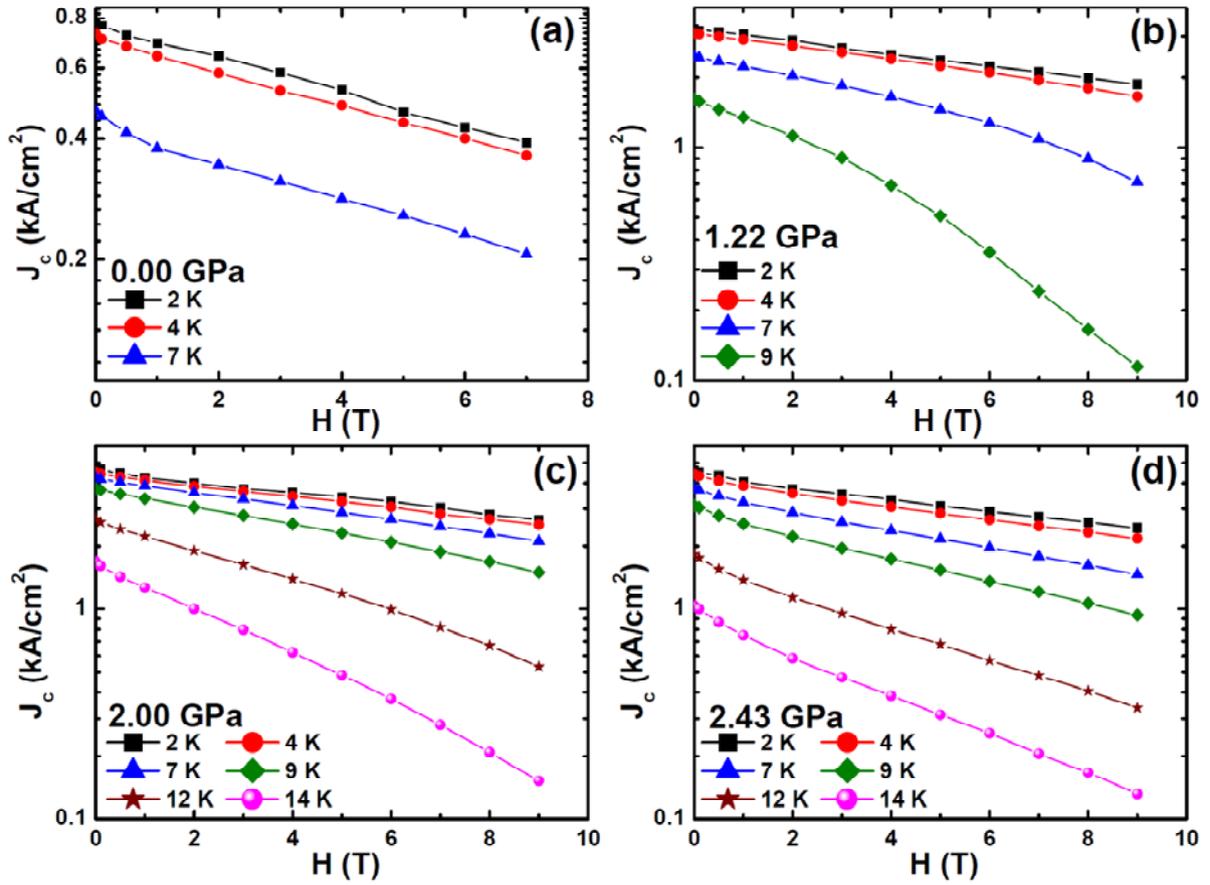

**Figure S2**. Depinning critical current density ($J_c$) of FeSe single crystals at pressures of (a) ambient, (b) 1.22 GPa, (c) 2.00 GPa, and (d) 2.43 GPa as a function of magnetic field. The weak-link behaviour that accompanies a sharp decrease in $J_c$ at low fields is not observed over the entire pressure range. Magnetic fields were applied parallel to the *ab*-plane of FeSe.

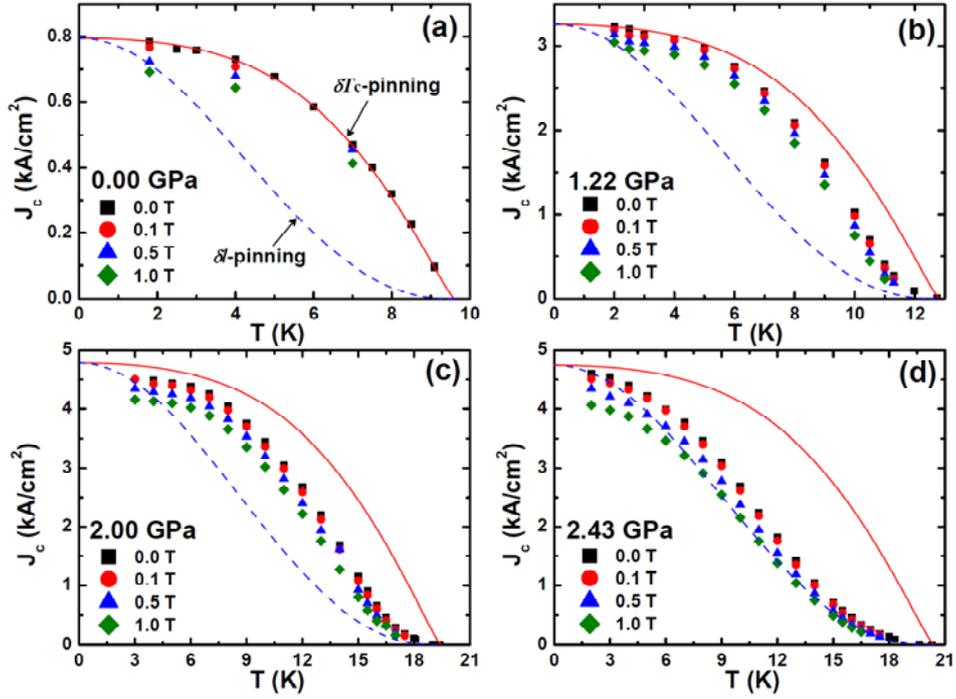

**Figure S3**. Temperature dependence of the depinning critical current density ($J_c$) under magnetic fields of 0 (squares), 0.1 (circles), 0.5 (triangles), and 1.0 Tesla (diamonds). The solid and the dashed lines are predictions from $\delta T_c$- and $\delta l$-pinning, respectively [S1,S2]. At ambient pressure, which is shown in panel (a), $J_c$ is well explained by $\delta T_c$-pinning and its overall behaviour is almost independent of magnetic field, indicating that spatial fluctuations in $T_c$ due to Se deficiencies and point defects are important to trapping vortices and that the pinning mechanism is insensitive to the field strength. When an antiferromagnetic phase is induced for pressures above 0.8 GPa [S3,S4], $J_c$ can no longer be explained by $\delta T_c$-pinning, rather it follows $\delta l$-pinning, suggesting that magnetic interactions between vortices and Fe moments are important to the pinning mechanism. Panels (b), (c), and (d) show $J_c$ at 1.22, 2.0, and 2.43 GPa, respectively. The two pinning mechanisms of $\delta T_c$-pinning and $\delta l$-pinning are related to the spatial fluctuations of the Ginzburg-Landau (GL) coefficient $\alpha$ associated with disorder in the superconducting transition temperature ($T_c$) and the spatial variations of the charge carrier mean free path ($l$) near lattice defects, respectively [S1,S2].

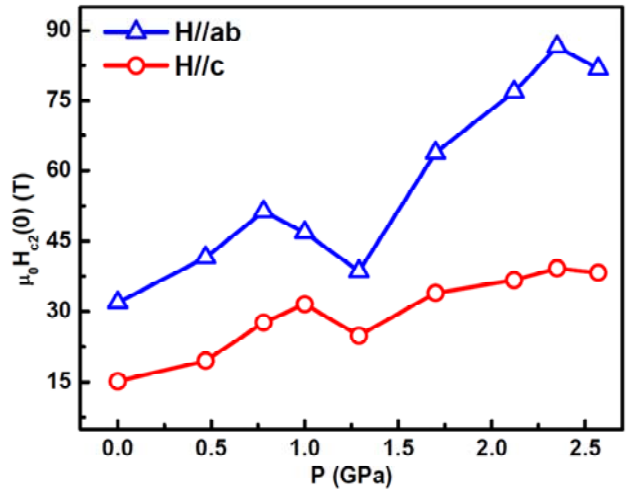

**Figure S4**. The zero temperature upper critical field, $\mu_0 H_{c2}(0)$, for FeSe as a function of applied pressure. $\mu_0 H_{c2}(0)$ is estimated from the Werthmaer Helfand and Hohenberg (WHH) model for magnetic field applied parallel (triangles) and perpendicular (circles) to the *ab*-plane [S5].

**Supplementary references**